\documentclass{jpsj-suppl}
\usepackage{txfonts} 
\usepackage{bm}

\newcommand{\bra}[1]{\langle \, #1 \, |}
\newcommand{\ket}[1]{| \, #1 \, \rangle}
\newcommand{\kket}[1]{\, #1 \, \rangle}

\newcommand{\PDG}{Agashe:2014kda}
\newcommand{\re}{\text{Re }}

\usepackage[normalem]{ulem}  
\usepackage[dvips]{color} 

\renewcommand\sout{\bgroup \color{red} \ULdepth=-.5ex \ULset}

\title{Compositeness of Hadrons and Near-Threshold Dynamics}

\author{Tetsuo \textsc{Hyodo}$^{1}$}

\inst{$^{1}$Yukawa Institute for Theoretical Physics, Kyoto University, Kyoto 606-8502, Japan }

\email{hyodo@yukawa.kyoto-u.ac.jp}

\recdate{October 5, 2015}

\abst{We present the recent developments in the studies of the structure of hadron resonances, focusing on the compositeness in terms of the hadronic degrees of freedom. We discuss the model dependence of the compositeness, and show that the structure of the near-threshold bound states and resonances is model-independently determined. The applications to various hadrons are summarized.}

\kword{Compositeness, hadronic molecules, low-energy universality, exotic hadrons}

\begin{document}
\maketitle

\section{Introduction}

One of the recent surprises in hadron physics is the observation of exotic hadrons in the heavy quark sectors~\cite{Belle:2011aa,Aaij:2015tga,\PDG}. These findings stimulate theoretical studies of the exotic composition of hadrons, such as multiquarks and hadronic molecular states. In fact, it is natural to expect the excitation of hadrons with a quark-antiquark pair creation in QCD, in addition to the internal excitation in the conventional constituent quark-model picture. In this sense, one may naively consider that a $N^{*}$ resonance is described by a superposition of all possible configurations,
\begin{equation}
   \ket{N^{*}} \stackrel{?}{=} N_{3q}\ket{uud}+N_{5q}\ket{uud\ q\bar{q}} + N_{\pi N}\ket{\pi N} + \dotsb .
   \label{eq:Nstar}
\end{equation}
The structure of the $N^{*}$ resonance may be understood  from the weights $N_{i}$ of the component $i$. However, it should be noted that there is no clear separation of each component in Eq.~\eqref{eq:Nstar}. One may define them in particular models, but there is no general definition in QCD. For a reasonable discussion on the structure of exotic hadrons, we must begin by establishing the suitable basis to characterize the structure of hadrons.

A promising approach is to consider the compositeness of hadrons which expresses the molecular nature of hadrons in terms of the asymptotic states in QCD. The compositeness was originally introduced to study the composite/elementary nature of particles using the field renormalization constant~\cite{Weinberg:1962hj,Weinberg:1965zz}. Later it turned out to be a useful quantity to characterize the structure of hadrons~\cite{Baru:2003qq,Hyodo:2011qc,Aceti:2012dd,Hyodo:2013iga,Sekihara:2014kya,Guo:2015daa,Kamiya:2015aea} (for a review, see Ref.~\cite{Hyodo:2013nka}). In this approach, the structure of the $N^{*}$ resonance is decomposed as
\begin{equation}
   \ket{N^{*}} =  \sqrt{X}\ket{MB} + \sqrt{Z}\ket{\text{others}} .
   \label{eq:Nstar2}
\end{equation}
where $X$ represents the compositeness of some meson-baryon channel $MB$. All the other components than $\ket{MB}$ are included in $\ket{\text{others}}$ which is regarded as the CDD pole contribution. As we will show below, various exotic hadrons have been studied from the viewpoint of the compositeness.

Here we present an introduction of the compositeness for the study of exotic hadrons, focusing on some subtle issues which are usually overlooked. We emphasize that the compositeness is in general a model-dependent quantity, like the wavefunction. On the other hand, the model-independent relation can be given in the weak-binding limit, thanks to the low-energy universality. Another important aspect is the interpretation of unstable particles for which the coefficients in Eq.~\eqref{eq:Nstar2} is in general complex. We show the recent discussion of the probabilistic interpretation of unstable particles.

\section{Formulation of compositeness for stable bound states}

We consider that some fundamental interaction (such as QCD) generates the physical system of one bound state coupled with one continuum scattering state in $s$ wave. We assume that there is no other coupled channels in the energy region considered. In this simplest system, we discuss the composite nature of the bound state.

The compositeness and elementariness are usually defined in the Hamiltonian system~\cite{Weinberg:1965zz,Sekihara:2014kya}. The essentially equivalent formulation can be given in the effective field theory (EFT) approach~\cite{Kamiya:2015aea} (see also Ref.~\cite{Chen:2013upa} for the use of EFT). We consider the nonrelativistic quantum field theory governed by the following Hamiltonian~\cite{Kaplan:1996nv,Braaten:2007nq}
\begin{align}
H&=H_{\mathrm{free}} + H_{\mathrm{int}} ,\quad \\
H_{\mathrm{free}} &=\int d\bm{r}
\biggl[\frac{1}{2 M} \mathbf{\nabla} \psi^\dagger(\bm{r}) \cdot\mathbf{\nabla} \psi(\bm{r}) +\frac{1}{2 m} \mathbf{\nabla} \phi^\dagger(\bm{r}) \cdot\mathbf{\nabla} \phi(\bm{r})  + \frac{1}{2 M_{0}} \mathbf{\nabla}  B_0^\dagger(\bm{r}) \cdot{\mathbf \nabla} B_0(\bm{r}) +  \nu_0 B_0^\dagger(\bm{r}) B_0(\bm{r})
\biggr] ,\\
H_{\mathrm{int}} &= \int d\bm{r}
\left[
g_{0} \left( B_0^\dagger(\bm{r}) \phi(\bm{r})\psi(\bm{r}) + \psi^\dagger(\bm{r})\phi^\dagger(\bm{r}) B_0(\bm{r}) \right) + \lambda_0 \psi^\dagger(\bm{r})\phi^\dagger(\bm{r}) \phi(\bm{r})\psi(\bm{r})
\right] .
\label{eq:Hint}
\end{align}
This theory consists of the fields $\psi$, $\phi$ and $B_{0}$. The $\psi\phi$ system couples to the $B_{0}$ state as well as interacts directly with themselves through the contact interactions. Because the statistics of the fields are not relevant for the two-body problem of the $\psi\phi$ system, we may simply regard them as bosons, $[\psi(\bm{r}),\psi^{\dag}(\bm{r}^{\prime})]=\delta^{3}(\bm{r}-\bm{r}^{\prime})$. As will be demonstrated below, this field theory provides an effective low-energy description of the bound state problem of our interest.

The vacuum of the system $\ket{0}$ is defined by the creation operators
\begin{equation}
   \tilde{\psi}(\bm{p})
   =\int d\bm{r}
   e^{-i\bm{p}\cdot\bm{r}}
   \psi(\bm{r}),
   \quad
   \tilde{\phi}(\bm{p})
   =\int d\bm{r}
   e^{-i\bm{p}\cdot\bm{r}}
   \phi(\bm{r}),
   \quad
   \tilde{B}_{0}(\bm{p})
   =\int d\bm{r}
   e^{-i\bm{p}\cdot\bm{r}}
   B_{0}(\bm{r}) ,
\end{equation}
as $\tilde{\psi}(\bm{p})\ket{0}=\tilde{\phi}(\bm{p})\ket{0}=\tilde{B}_{0}(\bm{p})\ket{0}=0 $. The eigenstates of the free Hamiltonian can be constructed from the vacuum $\ket{0}$ as
\begin{equation}
   \ket{B_{0}} = \frac{\tilde{B}_{0}^{\dag}(\bm{0})}{\sqrt{\mathcal{V}}}\ket{0},\quad
   \ket{\bm{p}} = \frac{\tilde{\psi}^{\dag}(\bm{p})
   \tilde{\phi}^{\dag}(-\bm{p})}{\sqrt{\mathcal{V}}}\ket{0} ,
\end{equation}
with the volume $\mathcal{V}=(2\pi)^{3}\delta^{3}(\bm{0})$. The eigenvalues are calculated as
\begin{equation}
   H_{\rm free}\ket{B_{0}} = \nu_{0}\ket{B_{0}},\quad
   H_{\rm free}\ket{\bm{p}}
   =\frac{p^{2}}{2\mu}\ket{\bm{p}} ,
\end{equation}
with the reduced mass $\mu=Mm/(M+m)$. These eigenstates satisfy the orthogonal conditions:
\begin{equation}
   \bra{B_{0}}\kket{B_{0}} =1 ,\quad
   \bra{\bm{p}}\kket{\bm{p}^{\prime}}
   =(2\pi)^{3}\delta^{3}(\bm{p}-\bm{p}^{\prime}),\quad
   \bra{B_{0}}\kket{\bm{p}}
   =0 .
\end{equation}

The phase symmetry of the interaction Hamiltonian indicates that the eigenstates of the full Hamiltonian $\ket{\Psi}$ of the $n_{\psi}+n_{B_{0}}=n_{\phi}+n_{B_{0}}=1$ sector can be given by
\begin{align}
   \ket{\Psi} 
   &=c\ket{B_{0}}+\int\frac{d\bm{p}}{(2\pi)^{3}} \chi(\bm{p})\ket{\bm{p}}
   \label{eq:twobody} ,
\end{align}
where $c$ and $\chi(\bm{p})$ represent the $B_{0}$ and $\bm{p}$ components of the wave function, respectively. The identity operator in this sector can be decomposed into the complete set as
\begin{equation}
   1 = \ket{B_{0}}\bra{B_{0}}+\int \frac{d\bm{p}}{(2\pi)^{3}}\ket{\bm{p}}\bra{\bm{p}}
   \label{eq:completeness} .
\end{equation}
We now consider that the system develops a bound state $\ket{B}$ as an eigenstate of the full Hamiltonian with the binding energy $B$, $H\ket{B} = -B\ket{B}$. The elementariness $Z$ and the compositeness $X$ of the bound state $\ket{B}$ are defined as
\begin{align}
Z &\equiv |\bra{B_{0}}\kket{B}|^{2}=|c|^{2},\quad 
X \equiv \int \frac{d\bm{p}}{(2\pi)^{3}} |\bra{\bm{p}}\kket{B}|^{2}
=\int \frac{d\bm{p}}{(2\pi)^{3}} |\chi(\bm{p})|^{2} .
\label{eq:normalization}
\end{align}
It follows from the normalization of the bound state wave function $\bra{B}\kket{B}=1$ and the completeness relation~\eqref{eq:completeness}  that the sum of $Z$ and $X$ is normalized to be unity:
\begin{equation}
   Z+X = 1, \quad Z,X\in [0,1] .
   \label{eq:boundstate}
\end{equation}
Namely, the elementariness and the compositeness are real and bounded. This is the necessary condition for the probabilistic interpretation of $Z$ and $X$. 

\section{Renormalization and model dependence}

In this way, the effective field theory can be used to define the bare states, orthogonality, and the completeness relation in the Hamiltonian system in Ref.~\cite{Sekihara:2014kya}. It is shown in Ref.~\cite{Sekihara:2014kya} that the elementariness and the compositeness can be related to the scattering observables. The forward scattering amplitude $f(E)$ of the $\psi\phi$ system can be written as
\begin{align}
   f(E) &= -\frac{\mu}{2\pi}t(E),\quad
   t(E) =\frac{1}{v(E)^{-1}-G(E)} ,
\end{align}
with
\begin{align}
   v(E) &=\lambda_0 + \frac{g_0^2}{E - \nu_0},
   \quad 
   G(E) = 
\int \frac{d\bm{p}}{(2\pi)^3}  \frac{1}{E-p^2/(2\mu)+i0^{+}} .
   \label{eq:vG}
\end{align}
Following Ref.~\cite{Sekihara:2014kya}, the elementariness and the compositeness can be expressed as
\begin{align}
   Z &= -g^{2}G^{2}(-B)v^{\prime}(-B),
   \quad
   X= -g^{2}G^{\prime}(-B)
   \label{eq:ZXscatt} ,
\end{align}
where 
\begin{align}
   v^{\prime}(E)=\frac{dv(E)}{dE} ,
   \quad 
   G^{\prime}(E)=\frac{dG(E)}{dE} ,
   \quad
   g^{2} =\lim_{E\to -B}(E+B)t(E) 
   = -\frac{1}{G^{2}(-B)v^{\prime}(-B)+G^{\prime}(-B)}.
\end{align}
The relation $Z+X=1$ with Eq.~\eqref{eq:ZXscatt} is called the sum rule, which can be derived by the generalized Ward identity~\cite{Sekihara:2010uz}, as well as by the general form of the scattering amplitude which satisfies unitarity and analyticity~\cite{Guo:2015daa}.

At this point, we note that the function $G(E)$ in Eq.~\eqref{eq:vG} has an ultraviolet divergence, because the interaction~\eqref{eq:Hint} is pointlike. In the EFT description, we are interested in the long wavelength dynamics of the system. In this case, the interaction can be regarded as a contact term, because the momentum is not large enough to resolve the microscopic details of the interaction. In other words, there is a momentum scale $\Lambda$ above which the detailed short-range structure of the interaction becomes relevant. The EFT description should be used only below this scale $\Lambda$. Regularizing the function $G(E)$ at this scale, we obtain
\begin{align}
   G(E;\Lambda) 
   = \frac{1}{2\pi^2} 
\int_{0}^{\Lambda} dp  \frac{p^{2}}{E-p^2/(2\mu)+i0^{+}} 
   &=-\frac{\mu}{\pi^{2}}\left(\Lambda-\sqrt{-2\mu E-i0^{+}}
   \arctan\frac{\Lambda}{\sqrt{-2\mu E-i0^{+}}}\right) .
   \label{eq:Greg}
\end{align}
The bare parameters $g_{0}$, $\nu_{0}$, $\lambda_0$ are determined at a given scale $\Lambda$ so as to reproduce the physical observables calculated by the scattering amplitude $f(E)$. 

Next we consider the variation of the cutoff scale $\Lambda$ (see a similar discussion in Ref.~\cite{Hyodo:2008xr}). If we choose a slightly shifted regularization scale $\Lambda+\delta\Lambda$, then the interaction $v(E)$ should be modified to reproduce the same scattering amplitude:
\begin{align}
   f(E) &= -\frac{\mu}{2\pi}\frac{1}{v_{\delta\Lambda}(E)^{-1}-G(E;\Lambda+\delta\Lambda)} .
\end{align}
Here the new interaction is given by
\begin{align}
   v_{\delta\Lambda}(E)
   &=\left(v(E)^{-1}
   +\frac{1}{2\pi^2} 
\int_{\Lambda}^{\Lambda+\delta\Lambda} dp  \frac{p^{2}}{E-p^2/(2\mu)+i0^{+}}  \right)^{-1} ,
\end{align}
where $v(E)$ is defined at scale $\Lambda$. In the expression of $Z$ and $X$ in Eq.~\eqref{eq:ZXscatt}, the residue of the bound state pole $g^{2}$ is invariant under the change of $\Lambda\to\Lambda+\delta\Lambda$:
\begin{align*}
   g^{2}
   &\to -\frac{1}{G^{2}(-B;\Lambda+\delta\Lambda)
   v_{\delta\Lambda}^{\prime}(-B)+G^{\prime}(-B;\Lambda+\delta\Lambda)} \\
   &= -\frac{1}{-G^{2}(-B;\Lambda+\delta\Lambda)
   v_{\delta\Lambda}(-B)^{2}[\{v^{-1}(-B)\}^{\prime}-G^{\prime}(-B;\Lambda)+G^{\prime}(-B;\Lambda+\delta\Lambda)]+G^{\prime}(-B;\Lambda+\delta\Lambda)} \\
   &= -\frac{1}{G^{2}(-B;\Lambda)v^{\prime}(-B)+G^{\prime}(-B;\Lambda)} 
\end{align*}
where we have used the bound state condition $v_{\delta\Lambda}(-B)G(-B;\Lambda+\delta\Lambda)=v(-B)G(-B;\Lambda)=1$. This shows that the residue is renormalization independent, as it should be. On the other hand, the other factors in Eq.~\eqref{eq:ZXscatt}, $G^{\prime}(-B;\Lambda)$ and $G^{2}(-B;\Lambda)v^{\prime}(-B)$, depend explicitly on $\delta\Lambda$. Hence, $Z$ and $X$ are in general renormalization-\textit{dependent} quantities. 

Let us consider the meaning of the regularization dependence of $Z$ and $X$. First of all, $Z$ is the field renormalization constant of the bare state $\ket{B_{0}}$ and hence  depends on the renormalization scale. The definition of the bare state $\ket{B_{0}}$ can be changed by the unitary transformation of the fields, which modifies the short range behavior of the interaction with keeping the observables unchanged. In the formulation of the Hamiltonian system, this dependence is related to the decomposition of the full Hamiltonian $H$ into $H_{\rm free}+H_{\rm int}$. The values of $Z$ and $X$ depend on how we define the bare state and the scattering states as the eigenstates of $H_{\rm free}$. In Eq.~\eqref{eq:Greg}, instead of the sharp cutoff for the regularization, one can also adopt the regulators of gaussian form, dipole form, and so on. All the different regularization schemes provide different values of $Z$ and $X$. This is equivalent to adopt the finite range interaction with different short range behavior. All these ambiguities are regarded as the model dependence of the compositeness. In this way, the values of $Z$ and $X$ in general reflect the microscopic details of the interaction.

\section{Weak-binding relations and low-energy universality}

Despite the model dependence of the compositeness in a general situation, model-independent discussion is possible in the weak binding limit. To appreciate this, we consider the small energy region satisfying $|E|\ll \Lambda^{2}/(2\mu)$. In this energy region, the function $G(E;\Lambda)$ can be approximated as
\begin{align}
   G(E;\Lambda) 
   &\to-\frac{\mu}{\pi^{2}}\left(\Lambda-\frac{\pi}{2}\sqrt{-2\mu E-i0^{+}}\right) .
   \label{eq:Gfnlow}
\end{align}
In this case, we find that $G^{\prime}(E)$ is independent of the cutoff $\Lambda$. Hence, $X=-g^{2}G^{\prime}(-B)=g^{2}\mu^{2}/(2\pi\sqrt{2\mu B})$ is scale \textit{independent}, if the binding energy satisfies $B\ll \Lambda^{2}/(2\mu)$. Because of the normalization~\eqref{eq:boundstate}, $Z$ is scale independent, which can also be shown with the relation $v_{\delta\Lambda}(E)\to [v(E)^{-1}+\mu\delta\Lambda/\pi^{2}]^{-1}$. In this way, $Z$ and $X$ of the weakly bound state are renormalization independent. We note that Eq.~\eqref{eq:Gfnlow} is the minimal expression of $G(E)$; the first term expresses the ultraviolet divergence and the second term is constrained by unitarity. The function $G(E)$ with any regulator will reduces to this form in the low-energy limit. In this sense, the compositeness is model independent in the weak binding limit. The expression of the compositeness with Eq.~\eqref{eq:Gfnlow} can be regarded as a nonrelativistic counterpart of the results obtained in the relativistic scattering amplitude satisfying unitarity and analyticity~\cite{Guo:2015daa}.

Because we consider the case with a small $B\ll \Lambda^{2}/(2\mu)$, various quantities can be expanded in powers of $B$. To perform a systematic expansion, we define the length scales $R$ and $R_{\rm typ}$ as 
\begin{align}
	R &= \frac{1}{\sqrt{2\mu B}},
	\quad R_{\rm typ}=\frac{1}{\Lambda} ,
\end{align}
where $R$ is determined by the tail of the bound state wave function and corresponds to the radius of the bound state. Because $\Lambda$ is the cutoff scale where the interaction is regarded as pointlike, $R_{\rm typ}$ is the largest length scale of the fundamental interaction. In the weak-binding limit where $R_{\rm typ}/R\ll 1$, the expansion of the scattering length $a_{0}=-f(0)$ leads to~\cite{Sekihara:2014kya,Kamiya:2015aea}
\begin{align}
	a_{0} &= R\left\{ \frac{2X}{1+X} + {\mathcal O}\left(\tfrac{R_{\mathrm{typ}}}{R}\right)\right\} , \label{eq:comp-rel-bound} 
\end{align}
This is the weak-binding relation firstly shown in Ref.~\cite{Weinberg:1965zz}. Because the coefficient of $\mathcal{O}(R)$ term is expressed by the compositeness $X$, the compositeness $X$ can be determined by the scattering length $a_{0}$ and the binding energy $B$ 
when the correction terms of order $\mathcal{O}(R_{\rm typ}/R)$ are neglected. 

In the scaling limit $R_{\rm typ}\to 0$, we obtain the relation $a_{0}=R$ which leads to $X=1$~\cite{Hyodo:2014bda,Hanhart:2014ssa}. This is a consequence of the low-energy universality; all the properties of the $s$-wave two-body scattering can be determined solely by the scattering length $a_{0}$ in the scaling limit. Equation~\eqref{eq:comp-rel-bound} shows that the leading violation of the scaling limit relation $a_{0}=R$ is determined by the compositeness $X$.

\section{Unstable particle and interpretation}

In the EFT formulation, it is possible to generalize the weak-binding relation to the unstable particles~\cite{Kamiya:2015aea}. Introducing the lower energy scattering channel into which the bound state decays, the compositeness $X$ of the quasi-bound state with the eigenenergy $E_{QB}$ is given by
\begin{align}
a_0
&=  R \Biggl\{\frac{2X}{1+X} + {\mathcal O}\left(\left|\tfrac{R_{\mathrm{typ}}}{R}\right| \right) + \sqrt{\frac{\mu^{\prime 3}}{\mu^{3}}} \mathcal{O} \left( \left| \tfrac{l}{R} \right|^{3}\right) \Biggr\}
\label{eq:comp-rel-quasi} ,
\quad
R=\frac{1}{\sqrt{-2\mu E_{QB}}},
\quad
l=\frac{1}{\sqrt{-2\mu \nu}} .
\end{align}
where $\nu$ is the energy difference of the thresholds of two scattering channels and $\mu^{\prime}$ is the reduced mass of the lower energy channel. If the correction terms of order ${\mathcal O}\left(\left|R_{\mathrm{typ}}/R\right| \right)$ and ${\mathcal O}\left(\left|l/R\right|^{3} \right)$ are neglected, the compositeness $X$ can be determined by the scattering length $a_{0}$ and the eigenenergy $E_{QB}$ model independently. In this case, the relation takes the same form with the bound state case~\eqref{eq:comp-rel-bound}, but now is valid among the complex values of $a_{0}$, $R$, and $X$. This is possible when the lower energy channel is sufficiently far away from the quasibound state to satisfy $|R|\gg l$, in addition to the weak binding condition $|R|\gg R_{\rm typ}$.

For unstable particles, the compositeness $X$ is given in general by a complex number~\cite{Hyodo:2013nka,Kamiya:2015aea}. In this case, the interpretation of the compositeness is an important problem. For an unstable state coupled with one scattering channel, the compositeness and the elementariness are given by
\begin{equation}
   Z+X = 1, \quad Z,X\in \mathbb{C} 
   \label{eq:unstable}
\end{equation}
In contrast to the stable bound state case~\eqref{eq:boundstate}, the complex $X$ is not bounded; it can take arbitrary number, provided that sum with $Z$ gives unity. A prescription proposed in Ref.~\cite{Kamiya:2015aea} is to define the real quantities as
\begin{align}
    \tilde{Z} \equiv &\frac{1 - |X| + |Z|}{2},
    \quad \tilde{X} \equiv \frac{1 - |Z| + |X|}{2},  
    \quad U \equiv |Z| +|X| -1 \label{eq:kaisyaku2-U} .
\end{align} 
We regard $U$ as the uncertainty of the determination~\cite{PLB73.389}, because it represents the degree of cancellation in the sum $Z+X$ which vanishes in the bound state case. If $U$ is small, we interpret $\tilde{Z}$ and $\tilde{X}$ are the probabilities of finding the bare (continuum) component in the quasi-bound state, because $\tilde{Z}$ and $\tilde{X}$ satisfy the normalization
\begin{align}
\tilde{Z}+\tilde{X} =1,\quad \tilde{Z},\tilde{X} \in [0,1] .
\label{eq:ZXQB}
\end{align}
This is reasonable because they reduce to the bound state relation in the $U\to 0$ limit. When $U$ is large, we should \textit{not} use $X$ as the measure of compositeness. In this case, we need to rely on other criteria such as the magnitude of the effective range~\cite{Baru:2003qq,Hyodo:2013iga}.

For the interpretation of the structure of unstable particles, several prescriptions have been proposed. In Ref.~\cite{Baru:2003qq}, the compositeness is defined through the integration of the spectral density, which can be obtained from the generalization of Eq.~\eqref{eq:ZXscatt}. Based on the property $\re Z+\re X = 1$, Ref.~\cite{Aceti:2014ala} argues that the real part of $Z$ and $X$ reflects the amount of the elementary and composite components. It is shown in Ref.~\cite{Guo:2015daa} that when the Laurent expansion of the scattering amplitude around the pole converges in a region of the physical axis, $X_{R} = |X|$ can be regarded as the probability of the compositeness. Although the results depend on the prescription, it should be noted that we obtain $\tilde{X}\sim \re X\sim |X|$ when $U$ is small. This means that all the prescriptions will give a consistent conclusion for a narrow width state.

\section{Compositeness of hadrons}

Here we summarize recent studies of compositeness of mesons (Tables~\ref{tbl:mesons}) and baryons (Tables~\ref{tbl:baryons}) in various approaches. We concentrate on the qualitative results of compositeness, irrespective of the main subject of the references. For details, one should refer to the original papers. 

\begin{table}[bp]
\caption{Summary of the compositeness of mesons.}
\label{tbl:mesons}
\begin{tabular}{lll} \hline
Meson & References & Result \\ \hline
$f_{0}(980)$ & \cite{Baru:2003qq}, \cite{Sekihara:2012xp}, \cite{Sekihara:2014qxa}, \cite{Sekihara:2014kya}, \cite{Kamiya:2015aea}, \cite{Guo:2015daa} & $\bar{K}K$ dominance \\
$a_{0}(980)$ & \cite{Baru:2003qq}, \cite{Sekihara:2012xp}, \cite{Sekihara:2014qxa}, \cite{Kamiya:2015aea} & small but non-negligible $\bar{K}K$  \\
$f_{0}(500)$ or $\sigma$ & \cite{Albaladejo:2012te}, \cite{Sekihara:2012xp}, \cite{Sekihara:2014kya}, \cite{Guo:2015daa} & $\pi\pi$ is not dominant \\
$K^{*}(800)$ or $\kappa$ & \cite{Sekihara:2014kya}, \cite{Guo:2015daa} & $K\pi$ dominance \\
$f_{0}(1710)$, $a_{0}(1450)$ & \cite{Guo:2015daa} & small two-pseudoscalar-meson components \\

$\rho(770)$ & \cite{Aceti:2012dd}, \cite{Sekihara:2014kya}, \cite{Guo:2015daa} & small $\pi\pi$ \\
$K^{*}(892)$ & \cite{Xiao:2012vv}, \cite{Sekihara:2014kya}, \cite{Guo:2015daa}  & small $K\pi$\\
$a_{1}(1260)$ & \cite{Guo:2015daa} & half $\rho\pi$ \\

$D_{s0}^*(2317)$ & \cite{Torres:2014vna}, \cite{Navarra:2015iea}, \cite{Guo:2015daa}  & $DK$ dominance \\
$D_{s1}^*(2460)$ & \cite{Torres:2014vna} & $D^{*}K$ dominance \\
$D_0^*(2400)$ & \cite{Navarra:2015iea}  & small $D\pi$, $D_{s}\bar{K}$ \\

$X(4260)$ & \cite{Guo:2015daa} & small $J/\psi f_{0}(500)$, $J/\psi f_{0}(980)$, $Z_{c}(3900)\pi$, $\omega\chi_{c0}$ \\

 \hline
\end{tabular}
\end{table}%

The evaluation of the compositeness of hadrons is initiated by Ref.~\cite{Baru:2003qq} for the study of $f_{0}(980)$ and $a_{0}(980)$ near the $\bar{K}K$ threshold. With the Flatte parameters determined by experiments, it is found that $f_{0}(980)$ and $a_{0}(980)$ have certain fraction of the $\bar{K}K$ component. The compositeness of $f_{0}(980)$ and $a_{0}(980)$ is evaluated in the leading order chiral unitary approach in Refs.~\cite{Sekihara:2012xp,Sekihara:2014qxa}. The results are qualitatively consistent with Ref.~\cite{Baru:2003qq}; the large amount of $f_{0}(980)$ is attributed to the $\bar{K}K$ component, while relatively small but non-negligible $\bar{K}K$ fraction is found for $a_{0}(980)$. Determinations of the pole position and the residues by more elaborated frameworks with the inverse amplitude method for the next-to-leading order chiral interaction~\cite{Sekihara:2014kya} and with the unitarized U(3) chiral perturbation theory~\cite{Guo:2015daa} support the $\bar{K}K$ molecular dominance of $f_{0}(980)$. On the other hand, the $\bar{K}K$ compositeness of $f_{0}(980)$ from the recent determinations of the Flatte parameters is rather scattered, as shown in Refs.~\cite{Sekihara:2014qxa,Kamiya:2015aea}. The results for $a_{0}(980)$ from the recent Flatte parameters indicate that only a small fraction is attributed to the $\bar{K}K$ component~\cite{Sekihara:2014qxa,Kamiya:2015aea}. To further clarify the nature of $f_{0}(980)$ and $a_{0}(980)$, precise determination of the near-threshold $\bar{K}K$ amplitude is mandatory.

For the lowest lying scaler meson $f_{0}(500)$ (or $\sigma$), it is pointed out that the $\pi\pi$ compositeness is small~\cite{Albaladejo:2012te}. The evaluations of the compositeness by the leading order chiral unitary approach~\cite{Sekihara:2012xp}, by the inverse amplitude method with the next-to-leading order chiral interaction~\cite{Sekihara:2014kya}, and by the unitarized U(3) chiral perturbation theory~\cite{Guo:2015daa} indicate the $\pi\pi$ component of $f_{0}(500)$ is not as large as the $\bar{K}K$ component of $f_{0}(980)$. The result for the strange $K^{*}(800)$ (or $\kappa$) is not conclusive in Ref.~\cite{Sekihara:2014kya} because of the large imaginary part, while the analysis in Ref.~\cite{Guo:2015daa} indicates a large $K\pi$ component. The two-meson components in higher scalar resonances, $f_{0}(1710)$ and $a_{0}(1450)$, are found to be small~\cite{Guo:2015daa}, in accordance with the interpretation of the glueball and the $\bar{q}q$ state, respectively.

The formulation for $p$-wave resonances is given in Ref.~\cite{Aceti:2012dd}. It is shown that the $\pi\pi$ molecule component in the $\rho(770)$ meson is found to be small. Consistent conclusions are obtained in more sophisticated approaches in Refs.~\cite{Sekihara:2014kya,Guo:2015daa}. The results for the $K^{*}$ is similar; the $K\pi$ component turns out to be small~\cite{Xiao:2012vv,Sekihara:2014kya,Guo:2015daa}. These results are in good agreement with the $\bar{q}q$ interpretation of the lowest vector mesons. The fraction of the $\rho\pi$ component in the axial vector $a_{1}(1260)$ meson is found to be about one half~\cite{Guo:2015daa}.

In the heavy sector, the amplitude constrained by the lattice data shows that $D_{s0}^*(2317)$ has large $KD$ component~\cite{Torres:2014vna}. The same conclusion is obtained in Refs.~\cite{Navarra:2015iea,Guo:2015daa} with the unitarized chiral perturbation theory. In the same manner, $D_{s1}^*(2460)$ is found to be dominated by the $KD^{*}$ component~\cite{Torres:2014vna}, and $D_0^*(2400)$ requires some contribution from the channels other than $D\pi+D_{s}\bar{K}$~\cite{Navarra:2015iea}. A related discussion on the lattice determination of the compositeness can be found in Ref.~\cite{Agadjanov:2014ana}. The compositeness of the hidden charm $X(4260)$ is calculated from the experimental analysis~\cite{Guo:2015daa}. The result indicates that the contributions from the nearby meson-meson channels are small.

\begin{table}[bp]
\caption{Summary of the compositeness of baryons.}
\label{tbl:baryons}
\begin{tabular}{lll} \hline
Baryon & References & Result \\ \hline
$\Lambda(1405)$, lower pole & \cite{Sekihara:2012xp}, \cite{Sekihara:2013sma}, \cite{Sekihara:2014kya}, \cite{Garcia-Recio:2015jsa}, \cite{Guo:2015daa} & $\pi\Sigma$ dominance \\
$\Lambda(1405)$, higher pole & \cite{Sekihara:2012xp}, \cite{Sekihara:2013sma}, \cite{Sekihara:2014kya}, \cite{Garcia-Recio:2015jsa}, \cite{Guo:2015daa}, \cite{Kamiya:2015aea} & $\bar{K}N$ dominance \\
$N(1535)$ & \cite{Sekihara:2014kya} & small meson-baryon components \\
$\Lambda(1670)$ & \cite{Sekihara:2014kya} & half $K\Xi$ \\
$\Xi(1690)$ & \cite{Sekihara:2015qqa} & $\bar{K}\Sigma$ dominance \\

$\Delta(1232)$, $\Sigma(1385)$,
$\Xi(1535)$, $\Omega$ & \cite{Aceti:2014ala}  & small meson-baryon components\\

$\Lambda(1520)$ & \cite{Aceti:2014wka}, \cite{Garcia-Recio:2015jsa}  & $d$-wave $\bar{K}N$ and $\pi\Sigma$ dominance \\

$\Lambda_{c}(2595)$ & \cite{Hyodo:2013iga},\cite{Garcia-Recio:2015jsa}, \cite{Guo:2015daa} & small $\pi\Sigma_{c}$ \\
$\Lambda_{c}(2625)$ & \cite{Garcia-Recio:2015jsa} & small meson-baryon components \\
$\Lambda_{b}(5912)$ & \cite{Garcia-Recio:2015jsa} & $\bar{B}^{*}N$ and $\pi\Sigma_{b}$ dominance \\
$\Lambda_{b}(5920)$ & \cite{Garcia-Recio:2015jsa} & $\pi\Sigma_{b}^{*}$ dominance \\
$P_{c}(4450)$ & \cite{Meissner:2015mza}  &  $\chi_{c1}N$ dominance \\

 \hline
\end{tabular}
\end{table}%

Turning to the baryonic sector, many works are dedicated to the $\Lambda(1405)$ resonance which is known to be associated with two resonance poles~\cite{Oller:2000fj,Jido:2003cb,Hyodo:2007jq,Hyodo:2011ur}. The compositeness of $\Lambda(1405)$ is evaluated by chiral unitary approach with the Weinberg-Tomozawa interaction~\cite{Sekihara:2012xp}, by the radiative decay~\cite{Sekihara:2013sma}, by the chiral approach with the next-to-leading order interaction~\cite{Sekihara:2014kya,Guo:2015daa,Kamiya:2015aea}, and by the SU(6) extension of the Weinberg-Tomozawa interaction~\cite{Garcia-Recio:2015jsa}. In all cases, it is found that the higher pole is dominated by the $\bar{K}N$ component, and the main component of the lower pole comes from the $\pi\Sigma$ state. The analysis in Ref.~\cite{Sekihara:2014kya} shows that a substantial component other than the meson-baryon channels is required for $N(1535)$, and the half of $\Lambda(1670)$ is made from the $K\Xi$ component. The structure of $\Xi(1690)$ is found to be dominated by the $\bar{K}\Sigma$ component~\cite{Sekihara:2015qqa}. 

The $p$-wave decuplet baryons are studied in Ref.~\cite{Aceti:2014ala}. The meson-baryon components are shown to be small, in accordance with the $qqq$ interpretation. The dominant component of $J^{P}=3/2^{-}$ $\Lambda(1520)$ is found to be $d$-wave $\bar{K}N$ and $\pi\Sigma$~\cite{Aceti:2014wka} and the $s$-wave channels give small contributions~\cite{Garcia-Recio:2015jsa}.

In the charmed baryon sector, the $\pi\Sigma_{c}$ component of $\Lambda_{c}(2595)$ is not the dominant one~\cite{Hyodo:2013iga}. Similar conclusion is drawn in the quantitative evaluations in Refs.~\cite{Garcia-Recio:2015jsa,Guo:2015daa} while the importance of the isospin breaking effect is pointed out~\cite{Guo:2015daa}. The meson-baryon components in $\Lambda_{c}(2625)$ is also found to be small~\cite{Garcia-Recio:2015jsa}. On the other hand, the negative parity $\Lambda_{b}$ states are dominated by the SU(6) meson-baryon channels~\cite{Garcia-Recio:2015jsa}.
The structure of the $P_{c}(4450)$ pentaquark is found to be a $\chi_{c1}p$ molecule~\cite{Meissner:2015mza}.

\section{Summary}

We present the recent studies of the structure of hadrons using the compositeness of hadrons. We demonstrate that the effective field theory provides the self-contained formulation for the discussion of the compositeness. Through the renormalization of EFT, we discuss the model dependence of the compositeness and its fate in the low-energy limit. The generalization to the unstable particles in various prescriptions is described.

There are many recent studies to evaluate the compositeness of exotic hadrons. We find that the conclusions are in good agreement with the expectation of other approaches, such as the $N_{c}$ scaling. This suggests that the examination of the compositeness is a powerful tool to unveil the internal structure of exotic hadron resonances. For further studies, it is important to establish the precise descriptions of the resonances in the two-body scattering amplitude.

This work is supported in part by JSPS KAKENHI Grants No. 24740152 and by the Yukawa International Program for Quark-Hadron Sciences (YIPQS).



\begin{thebibliography}{10}

\bibitem{Belle:2011aa}
Belle Collaboration, A.~Bondar {\em et~al.},
 Phys. Rev. Lett. {\bf 108} (2012) 122001.

\bibitem{Aaij:2015tga}
LHCb Collaboration, R.~Aaij {\em et~al.},
 Phys. Rev. Lett. {\bf 115} (2015) 072001.

\bibitem{Agashe:2014kda}
Particle Data Group, K.~Olive {\em et~al.},
 Chin. Phys. C {\bf 38} (2014) 090001.

\bibitem{Weinberg:1962hj}
S.~Weinberg,
 Phys. Rev. {\bf 130} (1963) 776.

\bibitem{Weinberg:1965zz}
S.~Weinberg,
 Phys. Rev. {\bf 137} (1965) B672.

\bibitem{Baru:2003qq}
V.~Baru, J.~Haidenbauer, C.~Hanhart, Y.~Kalashnikova and A.~E. Kudryavtsev,
 Phys. Lett. B {\bf 586} (2004)~53.

\bibitem{Hyodo:2011qc}
T.~Hyodo, D.~Jido and A.~Hosaka,
 Phys. Rev. C {\bf 85} (2012) 015201.

\bibitem{Aceti:2012dd}
F.~Aceti and E.~Oset,
 Phys. Rev. D {\bf 86} (2012) 014012.

\bibitem{Hyodo:2013iga}
T.~Hyodo,
 Phys. Rev. Lett. {\bf 111} (2013) 132002.

\bibitem{Sekihara:2014kya}
T.~Sekihara, T.~Hyodo and D.~Jido,
 PTEP {\bf 2015} (2015) 063D04.

\bibitem{Guo:2015daa}
Z.-H. Guo and J.~A. Oller,
arXiv:1508.06400 [hep-ph].

\bibitem{Kamiya:2015aea}
Y.~Kamiya and T.~Hyodo,
arXiv:1509.00146 [hep-ph]; in preparation.

\bibitem{Hyodo:2013nka}
T.~Hyodo,
 Int. J. Mod. Phys. A {\bf 28} (2013) 1330045.

\bibitem{Chen:2013upa}
G.-Y. Chen, W.-S. Huo and Q.~Zhao,
 Chin. Phys. C {\bf 39} (2015) 093101.

\bibitem{Kaplan:1996nv}
D.~B. Kaplan,
 Nucl. Phys. B {\bf 494} (1997) 471.

\bibitem{Braaten:2007nq}
E.~Braaten, M.~Kusunoki and D.~Zhang,
 Annals Phys. {\bf 323} (2008) 1770.

\bibitem{Sekihara:2010uz}
T.~Sekihara, T.~Hyodo and D.~Jido,
 Phys. Rev. C {\bf 83} (2011) 055202.

\bibitem{Hyodo:2008xr}
T.~Hyodo, D.~Jido and A.~Hosaka,
 Phys. Rev. C {\bf 78} (2008) 025203.

\bibitem{Hyodo:2014bda}
T.~Hyodo,
 Phys. Rev. C {\bf 90} (2014) 055208.

\bibitem{Hanhart:2014ssa}
C.~Hanhart, J.~Pelaez and G.~Rios,
 Phys. Lett. B {\bf 739} (2014) 375.

\bibitem{PLB73.389}
T.~Berggren,
 Phys. Lett. B {\bf 73} (1978) 389.

\bibitem{Aceti:2014ala}
F.~Aceti, L.~Dai, L.~Geng, E.~Oset and Y.~Zhang,
 Eur. Phys. J. A {\bf 50} (2014) 57.

\bibitem{Sekihara:2012xp}
T.~Sekihara and T.~Hyodo,
 Phys. Rev. C {\bf 87} (2013) 045202.

\bibitem{Sekihara:2014qxa}
T.~Sekihara and S.~Kumano,
 Phys. Rev. D {\bf 92} (2015) 034010.

\bibitem{Albaladejo:2012te}
M.~Albaladejo and J.~A. Oller,
 Phys. Rev. D {\bf 86} (2012) 034003.

\bibitem{Xiao:2012vv}
C.~Xiao, F.~Aceti and M.~Bayar,
 Eur. Phys. J. A {\bf 49} (2013) 22.

\bibitem{Torres:2014vna}
A.~Martinez~Torres, E.~Oset, S.~Prelovsek and A.~Ramos,
 JHEP {\bf 05} (2015) 153.

\bibitem{Navarra:2015iea}
F.~S. Navarra, M.~Nielsen, E.~Oset and T.~Sekihara,
 Phys. Rev. {\bf D92} (2015) 014031.

\bibitem{Agadjanov:2014ana}
D.~Agadjanov, F.~K. Guo, G.~Rios and A.~Rusetsky,
 JHEP {\bf 01} (2015) 118.

\bibitem{Sekihara:2013sma}
T.~Sekihara and S.~Kumano,
 Phys. Rev. C {\bf 89} (2014) 025202.

\bibitem{Garcia-Recio:2015jsa}
C.~Garcia-Recio, C.~Hidalgo-Duque, J.~Nieves, L.~L. Salcedo and L.~Tolos,
 Phys. Rev. D {\bf 92} (2015) 034011.

\bibitem{Sekihara:2015qqa}
T.~Sekihara,
 PTEP {\bf 2015} (2015) 091D01.

\bibitem{Aceti:2014wka}
F.~Aceti, E.~Oset and L.~Roca,
 Phys. Rev. C {\bf 90} (2014) 025208.

\bibitem{Meissner:2015mza}
U.-G. Meissner and J.~A. Oller,
Phys. Lett. B {\bf 751} (2015) 59.

\bibitem{Oller:2000fj}
J.~A. Oller and U.~G. Meissner,
 Phys. Lett. B {\bf 500} (2001) 263.

\bibitem{Jido:2003cb}
D.~Jido, J.~A. Oller, E.~Oset, A.~Ramos and U.~G. Meissner,
 Nucl. Phys. A {\bf 725} (2003) 181.

\bibitem{Hyodo:2007jq}
T.~Hyodo and W.~Weise,
 Phys. Rev. C {\bf 77} (2008) 035204.

\bibitem{Hyodo:2011ur}
T.~Hyodo and D.~Jido,
 Prog. Part. Nucl. Phys. {\bf 67} (2012) 55.

\end{thebibliography}

\end{document}